\documentclass[reprint,
superscriptaddress,
nofootinbib,
 amsmath,amssymb,
 aps,
pra,
]{revtex4-2}
\usepackage{graphicx}
\usepackage{amsmath, nccmath}
\usepackage{latexsym}
\usepackage{amsfonts}
\usepackage{amssymb}
\usepackage{array}
\usepackage{color}
\usepackage{subfigure}
\usepackage{xcolor}
\usepackage{verbatim}
\usepackage{gensymb}
\usepackage[bb=boondox]{mathalfa}
\usepackage{bigstrut}
\usepackage{multirow}
\usepackage{array}
\usepackage{tabularx}
\usepackage{booktabs}

\usepackage{comment}
\usepackage{physics}
\usepackage{float}

\newcommand{\myvar}{\mathrm{var}}

\usepackage[framemethod=TikZ]{mdframed}

\newenvironment{Frame1}[1][]{%
    \begin{mdframed}[%
        userdefinedwidth=\textwidth,
        frametitle={#1},
        skipabove=\baselineskip plus 2pt minus 1pt,
        skipbelow=\baselineskip plus 2pt minus 1pt,
        linewidth=0.5pt,
        frametitlerule=true,
        frametitlebackgroundcolor=gray!30,
        roundcorner=10pt,
        nobreak=true
    ]%
}{%
    \end{mdframed}
}

\newenvironment{Frame}[1][]{%
    \begin{mdframed}[%
        frametitle={#1},
        skipabove=\baselineskip plus 2pt minus 1pt,
        skipbelow=\baselineskip plus 2pt minus 1pt,
        linewidth=0.5pt,
        frametitlerule=true,
        frametitlebackgroundcolor=gray!30,
        roundcorner=10pt,
        nobreak=true
    ]%
}{%
    \end{mdframed}
}

\begin{document}
\newcommand{\Q}[1]{{\color{red}#1}}
\newcommand{\older}[1]{{\color{gray}#1}}
\newcommand{\red}[1]{{\color{red}#1}}
\newcommand{\green}[1]{{\color{green}#1}}
\newcommand{\update}[1]{{\color{magenta}#1}}
\newcommand{\Change}[1]{{\color{green}#1}}
\newcommand{\T}{\mathrm{Tr}}

\renewcommand{\thefootnote}{\fnsymbol{footnote}}

\title{Higher-order covariance matrices for non-Gaussian quantum states$\,^\dagger$}

\footnotetext[2]{This is the theory part of a forthcoming version of this paper, which will also report accompanying experimental results.} 

\renewcommand{\thefootnote}{\arabic{footnote}}

\author{Vojt\v{e}ch Kala}
\affiliation{Centre for Quantum Information and Communication, École polytechnique de Bruxelles, CP 165, Université libre de Bruxelles, 1050 Brussels, Belgium}
\author{Petr Marek}
\affiliation{Department of Optics, Palack\'y University, 17. listopadu 1192/12, 77146 Olomouc, Czech Republic}
\author{Nicolas J. Cerf}
\affiliation{Centre for Quantum Information and Communication, École polytechnique de Bruxelles, CP 165, Université libre de Bruxelles, 1050 Brussels, Belgium}

\date{\today}

\begin{abstract}
Covariance matrices lie at the heart of the powerful and well-established symplectic framework for describing continuous-variable Gaussian quantum states. However, since this framework  only relies on first- and second-order moments, it is not sufficient for the analysis of non-Gaussian states because their higher-order moments are essential to capture some of their key properties. Here, we define higher-order covariance matrices -- more precisely, covariance matrices built from higher-order quadrature monomials -- which provide a simple way to evaluate the effect of Gaussian transformations on non-Gaussian states and can be used, for example, to address nonlinear squeezing or non-Gaussian nullifiers. Higher-order covariance matrices can be estimated from homodyne measurement data using only a limited number of quadrature angles, which involves  matrices of moderate dimension compared with a full simulation in the Fock basis. The dimension of a higher-order covariance matrix does not depend on the span of the quantum states in Fock basis and, furthermore, scales only polynomially with the number of modes.
\end{abstract}

\maketitle

\section*{Introduction}
Quantum states with Gaussian quasiprobability distribution over the phase space are an essential part of quantum information and quantum optics. Together with linear transformations of quadratures preserving their Gaussianity, they form a versatile toolbox that appears in all applications arising from continuous-variable quantum information, from sensing \cite{Aasi2013} and cryptography \cite{Hajomer2024} to computation \cite{Konno,ZhengYu}. Gaussian states are fully characterized by their first and second moments, which give rise to a simple, yet practical and powerful formalism of covariance matrices \cite{Weedbrook,Brask}. Those fully describe Gaussian quantum states; even all linear transformations can be applied at the level of covariance matrices as symplectic transformations \cite{Weedbrook}.

Despite their wide applicability there are certain limitations to Gaussian states. When combined with Gaussian operations and measurements, they allow efficient classical simulation which rules them out as sufficient for quantum computation \cite{Mari2012}. This motivated large exploration of quantum states with non-Gaussian quasiprobability distributions which turned out to be a necessary resource for quantum computation \cite{Chabaud2023}. 

However, even though the Gaussian states and operations themselves are not sufficient for continuous variable quantum computing, they cannot be left out of the picture. First, when accompanied by a suitable non-Gaussian resource, they unlock universality \cite{Lloyd1999}. For example in the realm of Gottesman-Kitaev-Preskill states, they constitute the Clifford gates \cite{Gottesman2001}. Second, they allow for transformations of the non-Gaussian resources through the linear transformations of the phase space \cite{ZhengYu}. This ability can be used for transformation of one non-Gaussian resource into another, but as well needs to be taken into account when detecting non-Gaussianity. As the non-Gaussian resources are significantly more experimentally demanding compared to the Gaussian, Gaussian operations are considered free in the context of continuous variable quantum computing. Consequently, when witnessing a successful preparation of a non-Gaussian resource, we are interested in its presence, irrespective of its position or scaling in the phase space \cite{Kala2025,Provaznik2026}. This leads to definition of witnesses that are optimized over all Gaussian operations and simultaneously require a description that allows for Gaussian operations. 

Simultaneously, non-Gaussian states require characterization beyond the standard covariance matrix. Whereas for Gaussian states, all higher-order moments are trivial and can be obtained from the first two, for non-Gaussian states they deliver a new information. Therefore, other means of description were developed to deal with the non-Gaussian states, including analytical \cite{WalschaersPRX} and numerical methods \cite{Johansson2012}, which often become arduous, requiring matrices of high dimension dependent on the states and operations involved. Moreover, their scaling in the number of modes is exponential. Fortunately, it has been shown, that in some applications, the knowledge of higher moments of quadrature operators $x$ and $p$ ($[x,p]=i$) is sufficient; they can be used to recognize non-Gaussianity and quantify specific non-Gaussian resources in the continuous noise of quantum states \cite{Kala22,Provaznik2026}. Still, as those are usually defined in a form of a witness or phase sensitive, the requirement of ability to apply a general Gaussian transformation to the prepared state remains. Therefore, when detecting specific phase sensitive non-Gaussian properties, currently we either need a prior information of how the state is oriented in the phase space \cite{Kala2025P}, or we need to perform its full characterization. 

In order to solve this issue, we define higher-order covariance matrices. The formalism can be tailored to applications requiring limited knowledge of the non-Gaussian state. To great advantage, Gaussian operations can be applied by multiplication of small size matrices, allowing for efficient evaluation of non-Gaussian witnesses or simulation of decoherence. We illustrate our concept on two examples, the nonlinear squeezing \cite{Kala22} and recently defined nullifiers for non-Gaussian cluster states \cite{nonGnull}.

\section{Higher-order covariance matrix}
The usual measurement that is considered for non-Gaussian states in the context of continuous-variable quantum computing is the (phase-sensitive) homodyne measurement of the generalized quadrature
\begin{equation}\label{genq}
    X(\theta) = \cos(\theta)\, x + \sin(\theta) \, p,
\end{equation}
where $x$ and $p$ are the usual pair of canonically conjugate quadratures, with $[x,p]=i$. Performing this measurement for a set of quadrature angles $\{\theta_k\}$
enables the estimation of Weyl-symmetric monomials of quadrature operators via the expansion 
\begin{equation}\label{Wsym}
  :x^mp^n:_W  = \sum_{k=1}^{m+n} A_k \, X(\theta_k)^{m+n},
\end{equation}
where $\{A_k\}$ are real coefficients.
In Ref. \cite{nonGnull} it is shown, that the expansion always exists, although it is not necessarily unique.
Thus, higher-order moments and expectation values of nonlinear functions of quadratures can be estimated from measurements of the generalized quadratures \eqref{genq} measured under $m+n$ phase locks $\theta_k$.

Phase rotations play an important role, as can be illustrated with the simple example of Gaussian squeezing. Let us assume a squeezed vacuum with $\textrm{var}(x)$ being below the vacuum fluctuation level. If the state undergoes a $\pi/4$-rotation, its Gaussian squeezing can no longer be seen solely from measurements of the $x$ quadrature. Generally, when dealing with an arbitrary rotated squeezed vacuum, a minimization of $\textrm{var}(X(\theta))$ over $\theta$ is needed in order to unveil the presence of Gaussian squeezing.

A similar situation appears when dealing with non-Gaussian states, but with phase rotations being generalized by all Gaussian operations. Indeed, if the ``precious'' non-Gaussian resource is transformed by some Gaussian operation, the latter is viewed as irrelevant resource-wise since Gaussian operations are considered available and in principle can be reversed by an inverse operation. Therefore, non-Gaussian witnesses usually aim at identifying non-Gaussianity in the continuous quantum noise, regardless of displacement, rotation, and squeeze in phase space. In theory, this is represented by minimizing some functional over all Gaussian operations \cite{Kala22,Kala2025,Provaznik2026}, similarly to optimizing over rotation when evaluating Gaussian squeezing. However, to apply a Gaussian operation onto a non-Gaussian quantum state, we usually need the knowledge of its full density matrix, which makes the numerical calculation cumbersome.

As a way out, we show here that higher-order covariance matrices provide a formalism that easily incorporates rotations, as well as other Gaussian operations. The definition of higher-order covariance matrices is highly inspired by \cite{Kogias,Barrel,Tripathi}. Let us start by defining a set of Weyl-symmetric quadrature monomials
\begin{equation}\label{set}
    {\mathcal S}_N = \{ :x^n\, p^m :_W~|  n,m\in \mathbb N \text{~and~} 0\le n+m\le N  \} 
\end{equation}
where each monomial is of $N$th order at most. We require that ${\mathcal S}_N$ is closed under phase rotations, that is, every element of ${\mathcal S}_N$, when rotated by $R=\exp(-i\theta a^\dagger a)$, can be rewritten as a linear combination of elements of ${\mathcal S}_N$. We then build a vector $r$ containing all operators in ${\mathcal S}_N$. For example, with $N=2$, we may consider the five-dimensional vector (see Sec. \ref{sec:example})
\begin{equation}
r=    \begin{pmatrix}
x & p & x^2 & xp + px & p^2
\end{pmatrix}^T.
\end{equation}
Following the standard formulae, we define the higher-order mean vector, whose components are given by $\expval{r_i}$, and the higher-order covariance matrix, whose components are given by 
\begin{equation}\label{covnl}
\gamma_{ij} = \expval{\frac{r_ir_j+r_jr_i}{2}}-\expval{r_i}\expval{r_j} .
\end{equation}
Of course, we recover the usual mean vector and covariance matrix if we choose $r=    \begin{pmatrix}
x & p \end{pmatrix}^T$.

Interestingly, although $\expval{r_i}$ and $\gamma_{ij}$ go beyond the usual mean vector and covariance matrix, the way they transform under Gaussian operations (symplectic transformations)  remains formally unchanged. Given that the vector $r$ transforms under a Gaussian operation as
\begin{equation}\label{rtr}
    r'=Mr+v,
\end{equation}
where the matrix $M$ and vector $v$ can be found from the Heisenberg equations of motion, the higher-order mean vector and covariance matrix simply transform as
\begin{equation} 
    \expval{r'}=M\expval{r} + v
\end{equation}
and
\begin{equation}\label{transform}
    \gamma' = M\gamma M^T.
\end{equation}
For example, transformation \eqref{transform} can be proven from Eqs.~\eqref{covnl} and \eqref{rtr}: using Einstein summation convention, we have $\expval{r'_i}=M_{ik}\expval{r_k}+v_i$, and
\begin{equation}
    \begin{split}  
        \gamma_{ij}' =& \frac{1}{2}\langle (M_{ik}r_k+v_i)(M_{jl}r_l+v_j)+\\&(M_{jl}r_l+v_j)(M_{ik}r_k+v_i)\rangle-\\&\expval{M_{ik}r_k+v_i}\expval{M_{jl}r_l+v_j}\\
        =& \frac{1}{2}\expval{M_{ik}r_kr_lM_{jl}+M_{jl}r_lr_kM_{ik}}-M_{ik}\expval{r_k}\expval{r_l}M_{jl}\\
        =& M_{ik}\frac{1}{2}\expval{r_kr_l+r_lr_k}M_{jl}-M_{ik}\expval{r_k}\expval{r_l}M_{jl}\\
        =& M_{ik}(\frac{1}{2}\expval{r_kr_l+r_lr_k}-\expval{r_k}\expval{r_l})M_{lj}^T\\
        =& M_{ik} \gamma_{kl} M_{lj}^T .
    \end{split}
\end{equation}

Remarkably, the condition of physicality for higher-order covariance matrices keeps almost the usual form. Every physically acceptable $\gamma$ must obey \cite{Tripathi}
\begin{equation}\label{inqcon}
    \gamma+ \frac{i}{2}\expval{\Omega}\geq 0,
\end{equation}
where 
\begin{equation}
    i\, \Omega_{kl} = [r_k,r_l]
\end{equation}
The proof can be found in Appendix \ref{A:CondPhys}. Note that, in contrast with the usual covariance matrix formalism, the matrix $\Omega$ consists of operators and thus its mean value must be considered in Eq. \eqref{inqcon}. 

Overall, even though the higher-order covariance matrix formalism highly resembles the usual formalism for Gaussian states, its application is dramatically different and turns out to be very powerful for non-Gaussian states, as we illustrate in the next Section.

\section{Example with $N=2$}
\label{sec:example}

Now, we present a concrete example of a higher-order covariance matrix and derive the corresponding Gaussian transformations. In Section \ref{sec:application}, this example will be applied to evaluate the nonlinear squeezing and nullifiers for non-Gaussian cluster states. Even though they appear in different contexts and target distinct states, both applications can be approached based on the same higher-order covariance matrix, built from the five-dimensional vector of quadrature monomials (based on ${\mathcal S}_N$ with $N=2$)
\begin{equation}\label{vector}
r=    \begin{pmatrix}
x & p & x^2 & xp + px & p^2
\end{pmatrix}^T.
\end{equation}
As can be readily checked, it fulfills the requirement \eqref{set} on the set $\mathcal{S}_N$. From Eq. \eqref{covnl}, the higher-order covariance matrix can be written as
\begin{equation}\label{gamma}
\begin{medsize}
\begin{split}
    &\gamma=\\
&\begin{pmatrix}
\textrm{var}(x) & \textrm{cov}(x,p) & \textrm{cov}(x,x^2)& \textrm{cov}(x,xp+px)&\textrm{cov}(x,p^2)\\
. & \textrm{var}(p) & \textrm{cov}(p,x^2)& \textrm{cov}(p,xp+px)&\textrm{cov}(p,p^2)\\
.&.&\textrm{var}(x^2)& \textrm{cov}(x^2,xp+px)&\textrm{cov}(x^2,p^2)\\
.&.&.&\textrm{var}(xp+px)&\textrm{cov}(xp+px,p^2)\\
.&.&.&.&\textrm{var}(p^2)\\
\end{pmatrix}.
\end{split}   
\end{medsize}
\end{equation}

Eq.~\eqref{gamma} can be estimated from homodyne measurements under only six phase locks via Eq. \eqref{Wsym}, for example $\theta = -\frac{\pi}{4},0,\frac{\pi}{6},\frac{\pi}{4},\frac{\pi}{3},\frac{\pi}{2}$ (see Appendix \ref{A:Dict}). Note that the choice of $\theta$ is not unique. When estimated from experimental data, the physicality of $\gamma$ can be certified via Eq.~\eqref{inqcon}, where in this specific case
\begin{equation}
    \Omega=
\begin{pmatrix}
0&  1 &    0&  2x   & 2p\\
-1   &  0&  -2x   &  -2p&  0   \\
.&  .   &  0 &  4x^2   & 
2 (xp+px)\\
.   &  .&  .   &  0&  4p^2   \\
.&  .   & .&  .   &  0 
\end{pmatrix}.
\end{equation}

\subsection{Single-mode Gaussian transformations} 
Under a Gaussian operation, the higher-order covariance matrix can be transformed in the sense of Eq. \eqref{transform}.
The transformation matrix $M$ corresponding to a rotation, can be written as
\begin{widetext}
\begin{Frame1}[Rotation]
\begin{equation}\label{rotation}
    M_R(\theta)=
\begin{pmatrix}
\cos(\theta) & \sin(\theta) & 0& 0&0\\
-\sin(\theta) & \cos(\theta) & 0& 0&0\\
0& 0& \cos^2(\theta)&\cos(\theta)\sin(\theta)&\sin^2(\theta)\\
0& 0&-2\sin(\theta)\cos(\theta)   & \cos^2(\theta)-\sin^2(\theta)&2\sin(\theta)\cos(\theta)\\
0& 0&\sin^2(\theta)&-\sin(\theta)\cos(\theta)&\cos^2(\theta)
\end{pmatrix}
\end{equation}
\end{Frame1}
\end{widetext}


Similarly, the squeezing operation, acting as $x\rightarrow x/g$ and $p \rightarrow gp$, can be expressed via the diagonal matrix
\begin{Frame}[Gaussian squeezing]
   \begin{equation}\label{sq}
    M_S(g) = \begin{pmatrix}\frac{1}{g}&0&0&0&0\\0&g&0&0&0\\0&0&\frac{1}{g^2}&0&0\\0&0&0&1&0\\0&0&0&0&g^2\end{pmatrix}.
\end{equation} 
\end{Frame}

Finally, a displacement in $x$ corresponds to
\begin{Frame}[Displacement]
\begin{equation}
    M_D(d_x)=
\begin{pmatrix}
1 & 0 & 0& 0&0\\
0 & 1 & 0& 0&0\\
2d_x& 0& 1&0&0\\
0& 2d_x&0 & 1&0\\
0& 0&0&0&1
\end{pmatrix}
\end{equation}
\begin{equation}
    v = (d_x,0,d_x^2,0,0)^T,
\end{equation}
\end{Frame}
where, this time, $v$ is nonzero.

Note that it is not possible to include the cubic operation into the formalism as the set of operators constituting $r$ is not closed under the cubic operation. This is connected to the fact that cubic operation, together with rotation, can be used to approximate higher-order nonlinear operations \cite{Braunstein}.

We stress that the higher-order covariance matrix approach requires here the multiplication of only five-dimensional matrices, which can be quite advantageous compared to numerical simulations in Fock space. Such numerical simulations must be restricted to a finite-dimensional subspace of the Fock space that safely incorporates the whole quantum state in question. This truncation is therefore state dependent, and requires a good approximation of operators on this subspace \cite{Provaznik2022}.

\bigskip
\subsection{Multimode Gaussian operation}
An important part of the quantum optical description picture is a model of decoherence, especially losses. The usual approach is to introduce an auxiliary mode in a vacuum state which interacts with the mode of interest via beam splitter and is eventually traced out \cite{Kala22}. Therefore, we need to incorporate a two-mode Gaussian operation into our formalism, especially the beam splitter. For that purpose, the higher-order covariance matrix is extended to two modes, corresponding to quadratures $x,p$ and $x_0,p_0$. Compared to Eq. \eqref{vector}, the corresponding two-mode vector $r_2$ now contains monomials with all combinations (up to $N=2$), namely
\begin{equation}
r_2 = 
\begin{pmatrix}
     r& 2xx_0& 2pp_0& 2xp_0& 2x_0p& r_0   
\end{pmatrix}^T,
\end{equation}
where $r_0 = \begin{pmatrix} x_0& p_0 &x_0^2& x_0p_0+p_0x_0 &p_0^2\end{pmatrix}^T$. For $n$ modes, the vector $r_n$ must contain all  quadratic correlation terms, so that the dimension of the higher-order covariance matrix equals $5n + 2n(n-1)$, where the latter term is for increase in the correlations. This is in stark contrast with numerical simulations in a Fock subspace of dimension $dim$ which scales as $(dim^2)^n$ with the number of modes.

The matrix $M$ appearing in  transformation \eqref{rtr} for a beam splitter reads
\begin{widetext}
\begin{Frame1}[Beam splitter]
\begin{equation}\label{bs}
    M_{BS}(t)=
\begin{pmatrix}
t&  0&  0&   0&    0&    0&   0&   0&   0&    r&  0& 0&    0&    0\\
0&  t&  0&   0&    0&    0&   0&   0&   0&    0&  r& 0&    0&    0\\
0&  0&  t^2& 0&    0&    tr& 0&   0&   0&    0&  0& r^2& 0&    0\\
0&  0&  0&   t^2& 0&    0&   0&   tr& tr&  0&  0&  0& r^2&      0\\
0 &  0&  0&   0&    t^2 & 0&   tr& 0&   0&    0&  0&  0&    0&    r^2\\
-r & 0&  0&   0& 0 &    0 &   0&   0&   0&    t&  0&  0&    0&    0\\
0 &  -r& 0&   0&    0&    0&   0&   0&   0&    0&  t&  0&    0&    0\\
0 &  0&  r^2& 0&    0&   -tr& 0&   0&   0&    0&  0&  t^2& 0&    0\\
0 &  0&  0&   r^2& 0&   0&   0&  -tr& -tr& 0&  0&  0&    t^2& 0\\
    0 &  0&  0 &  0 & r^2 & 0 & -rt & 0 &0 & 0 & 0 & 0 & 0 & t^2
\end{pmatrix}.
\end{equation}
\end{Frame1}
\end{widetext} 

When the auxiliary mode with quadratures $x_0$ and $p_0$ is initially in the vacuum state, Eq. \eqref{bs} can be used to simulate pure losses. The single-mode higher-order covariance matrix after pure losses can be found in \eqref{aloss} in Appendix \ref{A:BS}. Compared to the Gaussian case, when the resulting covariance matrix can be written as a weighted sum of the initial and vacuum ones, the relation for a higher-order covariance matrix is slightly more complicated as it contains some cross terms; however, it still allows for a closed form.

\section{Applications}
\label{sec:application}

We now consider applications of the higher-order covariance matrix in the context of nonlinear squeezing and non-Gaussian cluster state nullifiers.

\subsection{Nonlinear squeezing} 
\label{subsec:IIIA}
One of the nonlinear functionals that can identify a non-Gaussian state consists of a variance of second order polynomial of quadrature operators, namely
\begin{equation}\label{var3}
    \textrm{var}(p+zx^2),
\end{equation}
where $z$ is a real number. The variance constitutes the cubic nonlinear squeezing \cite{Kala22,Brauer2021,Kala2025P,Konno_nlsq}. When its value falls below minimum over Gaussian states, it indicates non-Gaussianity of the considered state.
The variance can be estimated using \eqref{Wsym} from homodyne measurements with only four phase locks, for example $\theta = (0, \pi/2, \pi/4,-\pi/4)$~\cite{Moore19} 
\begin{equation}\label{var_4angles}
\begin{split}
\myvar_{\hat{\rho}} (\hat{p} + z \hat{x}^{2}) = \expval{X(\pi/2)^2} + z^2 \expval{X(0)^4} \\+ \frac{2\sqrt{2}z}{3}[\expval{X(\pi/4)^3}-\expval{X(-\pi/4)^3}] \\
- \frac{2z}{3}\expval{X(\pi/2)^3}- [\expval{X(\pi/2)} + z \expval{X(0)^2}]^2.
\end{split}
\end{equation}
Note that the orientation of the considered state in phase space must be known here.
The operator $p+zx^2$ is symmetric with respect to the transformation $x\rightarrow -x$. The same symmetry holds for states that minimize the variance \eqref{var3}, however it can be disrupted by an additional rotation of the state. Evaluation of the variance of the nonlinear operator in \eqref{var3} from homodyne measurements with just a few phase locks and simultaneously possibility to apply additional rotations and other Gaussian operations is provided via the higher-order covariance matrix formalism.

The chosen form \eqref{gamma} of the higher-order covariance matrix enables us to evaluate the variance \eqref{var3}, which is the main constituent of the nonlinear squeezing as
\begin{equation}\label{varY}
\textrm{var}(p+zx^2)= \begin{pmatrix}
0 & 1 & z & 0 & 0
\end{pmatrix}
\gamma
\begin{pmatrix}
0\\
1\\
z\\
0\\
0
\end{pmatrix}.
\end{equation}

As an example, we can consider the state
\begin{equation}
    \ket{\psi_3} = C(\chi)S(r)\ket{0},
\end{equation}
where $S(r) = \exp(r(a^2 - a^{\dagger 2})/2)$ and $C(\chi) = \exp(i\chi x^3/3)$. For parameters $r=-0.3$ and $\chi = 0.1$, we numerically simulate the state with the qutip library. The Fock subspace dimension was here chosen as $150$, to ensure that the last coefficients of the state are of the order $10^{-16}-10^{-17}$. The dimension of the Fock subspace must be carefully chosen and generally depends on the involved operations as well as their parameters, for example the strength of the squeezing. The corresponding higher-order covariance matrix is as follows

\begin{equation}
    \gamma_\psi=
\begin{pmatrix}
0.911 & 0.&  0.&  0.498 &  0. \\
0.  &  0.291 &  0.166&  0.  &  0.009 \\
0. &  0.166&  1.660&  0. & -0.409 \\
0.498&  0. &  0.  &  2.454&  0. \\
0.  &  0.009& -0.409&  0.  &  0.166
\end{pmatrix}
\end{equation}
and can be obtained either analytically, as the transformations in the state preparation have a known form in the Heisenberg picture, numerically in the Fock basis or from experimental homodyne data by means of \eqref{var_4angles}. Note that its dimension does not depend on the free parameters of the prepared state but on how the initial problem is defined. Furthermore, in the case of experimentally prepared state, the higher-order covariance matrix can be estimated from homodyne detection with just a few phase locks.

When the higher-order covariance matrix $\gamma$ is known, it enables estimation of the variance of the operator $p+zx^2$ \eqref{var3}. Importantly, Gaussian transformation of the state can be done on the higher-order covariance matrix via the transforming matrices, without the need to compute their Fock representations (here of dimension $150 \times 150$). 

Similarly, the formalism of higher-order covariance matrices simplifies the application of losses. In the discussed numerical simulation, addition of the auxiliary mode requires product of the Fock subspaces, requiring matrices of dimension $150^2 \times 150^2$ and an enough-accurate approximation of the beam splitter unitary operation of this dimension. The higher-order covariance matrix incorporates losses with fourteen dimensional matrices or as a single mode transformation of a matrix of dimension five (see Appendix \ref{A:BS}).

In order to illustrate the improvement obtained by using the higher-order covariance matrix formalism, we can compare the value of \eqref{var3} as obtained with the higher-order covariance matrix and its Gaussian limit. The variance \eqref{var3} can be rewritten as $var(p+zx^2)=var(p)+2zcov(p,x^2)+z^2var(x^2)$. Assuming a Gaussian distribution, it holds that
\begin{equation}
    \begin{split}
        \textrm{cov}(p,x^2) &= 2\expval{x}\textrm{cov}(x,p)\\
        \textrm{var}(x^2) &= 2\textrm{var}(x)^2 + 4\expval{x}^2\textrm{var}(x).
    \end{split}
\end{equation}

\begin{figure}
\includegraphics[width=1\columnwidth]{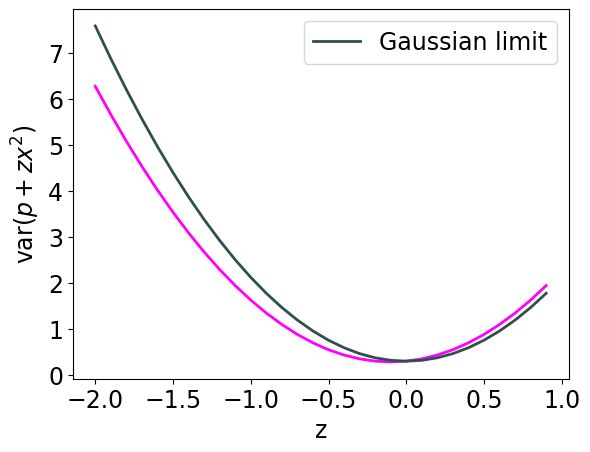}
\caption{Variance of the operator $p+zx^2$ (magenta) as a function of $z$, computed with the higher-order covariance matrix and as its Gaussian limit.}
\label{rot}
\end{figure}

The comparison is shown in Fig. \ref{rot}. The two values coincide for $z=0$, however when the absolute value of $z$ increases, the value of the Gaussian limit starts to significantly differ from the actual value.

\subsection{Non-Gaussian cluster state nullifier}
Photon subtracted Gaussian squeezed states can be found as a ground state of the operator
\begin{equation}\label{nullifier}
\begin{split}
        &S(r)OS^\dagger(r) = S(r)(n-1)^2S^\dagger(r)\\
        &= S(r)(x^4 + p^4 + x^2p^2 + p^2x^2 - 2(x^2 + p^2))S^\dagger(r).
\end{split}
\end{equation}
Therefore, using linearity of the beam splitter interaction, this operator can be used for definition of nullifiers for non-Gaussian cluster states prepared from photon subtracted squeezed states \cite{nonGnull}.

In contrast with Sec. \ref {subsec:IIIA}, we are now interested in the mean value of $O$, not its variance. A matrix containing the required moments can be easily obtained from the covariance matrix and vector of mean values as
\begin{equation}
    W = \gamma + \expval{r}\otimes \expval{r}^T.
\end{equation}
Then, the expectation value of the operator $O$ \eqref{nullifier} is given by
\begin{equation}
    \expval{O} = u_1^TWu_1 + u_2^TWu_2 + u_3^TWu_3,
\end{equation}
with
\begin{equation}
    \begin{split}
        u_1 =& (-2,0,0,0,0)^T\\
        u_2 = & (0,-2,0,0,0)^T\\
        u_3 = & (0,0,1,0,1)^T.
    \end{split}
\end{equation}
The moment matrix $W$ transforms in the same manner as the higher-order covariance matrix $\gamma$ under Gaussian transformations with $v=0$ in Eq. \eqref{rtr}. Thus, the squeezing operation in Eq. \eqref{nullifier} can be applied via Eqs. \eqref{sq} and \eqref{transform}.

Under a general Gaussian transformation in the form of \eqref{rtr}, the moment matrix $W$ transforms as
\begin{equation}
    W' = MWM^T + M\expval{r}\otimes \expval{v}^T + \expval{v}\otimes \expval{r}^TM^T + \expval{v}\otimes \expval{v}^T.
\end{equation}

For the description of $n$-mode cluster state, a $5n + 2n(n-1)$-dimensional higher-order covariance matrix is needed, which enables the application of all single-mode Gaussian operations and beam splitters.

\section{Conclusion}
The formalism of higher-order covariance matrices significantly simplifies description and transformations of the information about non-Gaussian states needed for witnessing non-Gaussianity. We presented two examples that utilize a higher-order covariance matrix of dimension five. We showed the extension beyond a single mode, which can be used for simulation of decoherence or description of a cluster state. The higher-order covariance matrices can be directly estimated from a homodyne measurement, which is ubiquitous in continuous-variable quantum information and computation.

\section*{Acknowledgments}
VK acknowledges support from the Fonds de la Recherche Scientifique (F.R.S.-FNRS) under project CHEQS within the Excellence of Science (EOS) program.
PM acknowledges the financial support
of the Czech Science Foundation (project 25-17472S).
PM also acknowledges European Union’s HORIZON Research and Innovation
Actions under Grant Agreement no. 101080173 (CLUSTEC).
PM acknowledges
a grant from the Programme Johannes Amos Comenius
under the Ministry of Education,Youth and Sports of the Czech
Republic reg. no. CZ.02.01.01/00/22\textunderscore 008/0004649. 
NJC acknowledges support from the Fonds de la Recherche Scientifique (F.R.S.-FNRS) under Grant No T.0060.26.

\bibliographystyle{unsrtnat}
\bibliography{references}

@article{Brask,
Author = {Jonatan Bohr Brask},
Title = {Gaussian states and operations -- a quick reference},
Year = {2021},
Eprint = {arXiv:2102.05748},
journal = {arXiv:2102.05748 [quant-ph]}
}

@article{Gottesman2001,
  title = {Encoding a qubit in an oscillator},
  author = {Gottesman, Daniel and Kitaev, Alexei and Preskill, John},
  journal = {Phys. Rev. A},
  volume = {64},
  issue = {1},
  pages = {012310},
  numpages = {21},
  year = {2001},
  month = {Jun},
  publisher = {American Physical Society},
  doi = {10.1103/PhysRevA.64.012310},
  url = {https://link.aps.org/doi/10.1103/PhysRevA.64.012310}
}

@article{Moore19,
	doi = {10.1088/1367-2630/ab5690},
	url = {https://doi.org/10.1088/1367-2630/ab5690},
	year = {2019},
	month = {nov},
	publisher = {{IOP} Publishing},
	volume = {21},
	number = {11},
	pages = {113050},
	author = {Darren W Moore and Andrey A Rakhubovsky and Radim Filip},
	title = {Estimation of squeezing in a nonlinear quadrature of a mechanical oscillator},
	journal = {New Journal of Physics}
}

@article{Brauer2021,
author = {\v{S}imon Br\"{a}uer and Petr Marek},
journal = {Opt. Express},
number = {14},
pages = {22648--22658},
publisher = {OSA},
title = {Generation of quantum states with nonlinear squeezing by Kerr nonlinearity},
volume = {29},
month = {Jul},
year = {2021},
url = {http://www.opticsexpress.org/abstract.cfm?URI=oe-29-14-22648},
doi = {10.1364/OE.427637},
}

@article{Aasi2013,
    author = {Aasi, J. and Abadie, J. and Abbott, B. and others},
    title = {Enhanced sensitivity of the LIGO gravitational wave detector by using squeezed states of light},
	volume = {7},
	pages = {613-619},
	year = {2013},
	doi = {10.1038/nphoton.2013.177},
	journal = {Nat. Phot.}
}

@article{Lloyd1999,
    author = {Seth Lloyd and Samuel L. Braunstein},
    title = {Quantum Computation over Continuous Variables},
  journal = {Phys. Rev. Lett.},
  volume = {82},
  issue = {8},
  pages = {1784--1787},
  numpages = {0},
  year = {1999},
  month = {Feb},
  publisher = {American Physical Society},
  doi = {10.1103/PhysRevLett.82.1784},
  url = {https://link.aps.org/doi/10.1103/PhysRevLett.82.1784}
}

@article{Provaznik2022,
  title = {Taming numerical errors in simulations of continuous variable non-Gaussian state preparation},
  volume = {12},
  ISSN = {2045-2322},
  url = {http://dx.doi.org/10.1038/s41598-022-19506-9},
  DOI = {10.1038/s41598-022-19506-9},
  number = {1},
  journal = {Scientific Reports},
  publisher = {Springer Science and Business Media LLC},
  author = {Provazník,  Jan and Filip,  Radim and Marek,  Petr},
  year = {2022},
  month = Oct 
}

@article{Provaznik2026,
Author = {Jan Provazník and Šimon Bräuer and Vojtěch Kala and Jaromír Fiurášek and Petr Marek},
Title = {Witnesses of non-Gaussian features as lower bounds of stellar rank},
Year = {2026},
Eprint = {arXiv:2603.03185},
journal = {arXiv:2603.03185 [quant-ph]}
}

@article{nonGnull,
  doi = {10.48550/ARXIV.2505.21066},
  url = {https://arxiv.org/abs/2505.21066},
  author = {Kala,  Vojtěch and Breum,  Casper A. and Larsen,  Mikkel V. and Andersen,  Ulrik L. and Neergaard-Nielsen,  Jonas S. and Filip,  Radim and Marek,  Petr},
  title = {Nullifiers of non-Gaussian cluster states through homodyne measurement},
  journal = {arXiv:2505.21066 [quant-ph]},
  year = {2025},
  copyright = {arXiv.org perpetual,  non-exclusive license}
}

@article{Kala22,
author = {Vojt\v{e}ch Kala and Radim Filip and Petr Marek},
journal = {Opt. Express},
number = {17},
pages = {31456--31471},
publisher = {Optica Publishing Group},
title = {Cubic nonlinear squeezing and its decoherence},
volume = {30},
month = {Aug},
year = {2022},
url = {https://opg.optica.org/oe/abstract.cfm?URI=oe-30-17-31456},
doi = {10.1364/OE.464759},
}

@article{Johansson2012,
  title = {QuTiP: An open-source Python framework for the dynamics of open quantum systems},
  author = {Johansson,  J.R. and Nation,  P.D. and Nori,  Franco},
  volume = {183},
  ISSN = {0010-4655},
  url = {http://dx.doi.org/10.1016/j.cpc.2012.02.021},
  DOI = {10.1016/j.cpc.2012.02.021},
  number = {8},
  journal = {Computer Physics Communications},
  publisher = {Elsevier BV},
  year = {2012},
  month = Aug,
  pages = {1760–1772}
}

@article{WalschaersPRX,
  title = {Practical Framework for Conditional Non-Gaussian Quantum State Preparation},
  author = {Walschaers, Mattia and Parigi, Valentina and Treps, Nicolas},
  journal = {PRX Quantum},
  volume = {1},
  issue = {2},
  pages = {020305},
  numpages = {14},
  year = {2020},
  month = {Oct},
  publisher = {American Physical Society},
  doi = {10.1103/PRXQuantum.1.020305},
  url = {https://link.aps.org/doi/10.1103/PRXQuantum.1.020305}
}

@article{Chabaud2023,
  title = {Resources for Bosonic Quantum Computational Advantage},
  author = {Chabaud, Ulysse and Walschaers, Mattia},
  journal = {Phys. Rev. Lett.},
  volume = {130},
  issue = {9},
  pages = {090602},
  numpages = {7},
  year = {2023},
  month = {Mar},
  publisher = {American Physical Society},
  doi = {10.1103/PhysRevLett.130.090602},
  url = {https://link.aps.org/doi/10.1103/PhysRevLett.130.090602}
}

@article{Mari2012,
    author = {Andrea Mari and Jens Eisert},
    title = {Positive Wigner Functions Render Classical Simulation of Quantum Computation Efficient},
  journal = {Phys. Rev. Lett.},
  volume = {109},
  issue = {23},
  pages = {230503},
  numpages = {5},
  year = {2012},
  month = {Dec},
  publisher = {American Physical Society},
  doi = {10.1103/PhysRevLett.109.230503},
  url = {https://link.aps.org/doi/10.1103/PhysRevLett.109.230503}
}

@article{Weedbrook,
  title = {Gaussian quantum information},
  author = {Weedbrook, Christian and Pirandola, Stefano and Garc\'{\i}a-Patr\'on, Ra\'ul and Cerf, Nicolas J. and Ralph, Timothy C. and Shapiro, Jeffrey H. and Lloyd, Seth},
  journal = {Rev. Mod. Phys.},
  volume = {84},
  issue = {2},
  pages = {621--669},
  numpages = {0},
  year = {2012},
  month = {May},
  publisher = {American Physical Society},
  doi = {10.1103/RevModPhys.84.621},
  url = {https://link.aps.org/doi/10.1103/RevModPhys.84.621}
}

@article{Kala2025P,
  title = {Nonlinear squeezing generation via multimode PDC and single photon measurement},
  volume = {33},
  ISSN = {1094-4087},
  url = {http://dx.doi.org/10.1364/OE.550358},
  DOI = {10.1364/oe.550358},
  number = {6},
  journal = {Optics Express},
  publisher = {Optica Publishing Group},
  author = {Kala,  Vojtěch and Kopylov,  Denis and Marek,  Petr and Sharapova,  Polina},
  year = {2025},
  month = Mar,
  pages = {14000}
}

@article{Kala2025,
Author = {Vojtěch Kala and Jiří Fadrný and Michal Neset and Jan Bílek and Petr Marek and Miroslav Ježek},
Title = {Genuine Continuous Quantumness},
Year = {2025},
Eprint = {arXiv:2503.07574},
journal = {arXiv:2503.07574 [quant-ph]}
}

@article{ZhengYu,
  title = {Gaussian Conversion Protocols for Cubic Phase State Generation},
  author = {Zheng, Yu and Hahn, Oliver and Stadler, Pascal and Holmvall, Patric and Quijandr\'{\i}a, Fernando and Ferraro, Alessandro and Ferrini, Giulia},
  journal = {PRX Quantum},
  volume = {2},
  issue = {1},
  pages = {010327},
  numpages = {25},
  year = {2021},
  month = {Feb},
  publisher = {American Physical Society},
  doi = {10.1103/PRXQuantum.2.010327},
  url = {https://link.aps.org/doi/10.1103/PRXQuantum.2.010327}
}

@article{Konno,
  title = {Non-Clifford gate on optical qubits by nonlinear feedforward},
  author = {Konno, Shunya and Asavanant, Warit and Fukui, Kosuke and Sakaguchi, Atsushi and Hanamura, Fumiya and Marek, Petr and Filip, Radim and Yoshikawa, Jun-ichi and Furusawa, Akira},
  journal = {Phys. Rev. Res.},
  volume = {3},
  issue = {4},
  pages = {043026},
  numpages = {11},
  year = {2021},
  month = {Oct},
  publisher = {American Physical Society},
  doi = {10.1103/PhysRevResearch.3.043026},
  url = {https://link.aps.org/doi/10.1103/PhysRevResearch.3.043026}
}

@article{Hajomer2024,
  title = {Continuous-variable quantum passive optical network},
  volume = {13},
  ISSN = {2047-7538},
  url = {http://dx.doi.org/10.1038/s41377-024-01633-9},
  DOI = {10.1038/s41377-024-01633-9},
  number = {1},
  journal = {Light Sci. Appl.},
  publisher = {Springer Science and Business Media LLC},
  author = {Hajomer,  Adnan A. E. and Derkach,  Ivan and Filip,  Radim and Andersen,  Ulrik L. and C. Usenko,  Vladyslav and Gehring,  Tobias},
  year = {2024},
  month = {Oct} 
}

@article{Braunstein,
  title = {Quantum information with continuous variables},
  author = {Braunstein, Samuel L. and van Loock, Peter},
  journal = {Rev. Mod. Phys.},
  volume = {77},
  issue = {2},
  pages = {513--577},
  numpages = {0},
  year = {2005},
  month = {Jun},
  publisher = {American Physical Society},
  doi = {10.1103/RevModPhys.77.513},
  url = {https://link.aps.org/doi/10.1103/RevModPhys.77.513}
}

@article{Tripathi,
doi = {10.1088/1367-2630/ab9ce7},
url = {https://dx.doi.org/10.1088/1367-2630/ab9ce7},
year = {2020},
month = {jul},
publisher = {IOP Publishing},
volume = {22},
number = {7},
pages = {073055},
author = {Vinay Tripathi and Chandrashekar Radhakrishnan and Tim Byrnes},
title = {Covariance matrix entanglement criterion for an arbitrary set of operators},
journal = {New Journal of Physics}
}

@article{Barrel,
  title = {Hierarchy of Nonlinear Entanglement Dynamics for Continuous Variables},
  author = {Zhang, Da and Barral, David and Cai, Yin and Zhang, Yanpeng and Xiao, Min and Bencheikh, Kamel},
  journal = {Phys. Rev. Lett.},
  volume = {127},
  issue = {15},
  pages = {150502},
  numpages = {6},
  year = {2021},
  month = {Oct},
  publisher = {American Physical Society},
  doi = {10.1103/PhysRevLett.127.150502},
  url = {https://link.aps.org/doi/10.1103/PhysRevLett.127.150502}
}

@article{Kogias,
  title = {Hierarchy of Steering Criteria Based on Moments for All Bipartite Quantum Systems},
  author = {Kogias, Ioannis and Skrzypczyk, Paul and Cavalcanti, Daniel and Ac\'{\i}n, Antonio and Adesso, Gerardo},
  journal = {Phys. Rev. Lett.},
  volume = {115},
  issue = {21},
  pages = {210401},
  numpages = {7},
  year = {2015},
  month = {Nov},
  publisher = {American Physical Society},
  doi = {10.1103/PhysRevLett.115.210401},
  url = {https://link.aps.org/doi/10.1103/PhysRevLett.115.210401}
}

@article{Konno_nlsq,
  title = {Nonlinear Squeezing for Measurement-Based Non-Gaussian Operations in Time Domain},
  author = {Konno, Shunya and Sakaguchi, Atsushi and Asavanant, Warit and Ogawa, Hisashi and Kobayashi, Masaya and Marek, Petr and Filip, Radim and Yoshikawa, Jun-ichi and Furusawa, Akira},
  journal = {Phys. Rev. Applied},
  volume = {15},
  issue = {2},
  pages = {024024},
  numpages = {9},
  year = {2021},
  month = {Feb},
  publisher = {American Physical Society},
  doi = {10.1103/PhysRevApplied.15.024024},
  url = {https://link.aps.org/doi/10.1103/PhysRevApplied.15.024024}
}

\subsection{Physicality condition on $\gamma$}\label{A:CondPhys}
It is important to ensure, that the higher order covariance matrix corresponds to a physical state. This can be done in the following way.
For the matrix $\gamma$ \eqref{covnl} holds
\begin{equation}
    \begin{split}
    \gamma_{ij} =& \expval{\frac{r_ir_j+r_jr_i}{2}}-\expval{r_i}\expval{r_j}\\
    =& \expval{\frac{2r_ir_j+[r_i,r_j]}{2}}-\expval{r_i}\expval{r_j}\\
    =& \expval{r_ir_j}-\expval{r_i}\expval{r_j}-\expval{\frac{[r_i,r_j]}{2}}
    \end{split}
\end{equation}
from that follows:
\begin{equation}\label{comm}
    \begin{split}
    \gamma_{ij}+ \expval{\frac{[r_i,r_j]}{2}}=& \expval{r_ir_j}-\expval{r_i}\expval{r_j}\\
    =&\expval{(r_i-\expval{r_i})(r_j-\expval{r_j})}.
    \end{split}
\end{equation}
We can define a matrix
\begin{equation}
    i\Omega = [r_i,r_j]
\end{equation}
and write \eqref{comm} as
\begin{equation}\label{rovnice}
    \gamma_{ij}+ \frac{i}{2}\expval{\Omega}_{ij}=
    \expval{(r_i-\expval{r_i})(r_j-\expval{r_j})}.
\end{equation}
Now, the term $ \expval{(r_i-\expval{r_i})(r_j-\expval{r_j})}$ is positive semidefinite \cite{Tripathi}, so is the left hand side of \eqref{rovnice}.

\subsection{Effect of optical losses}\label{A:BS}

It is possible to analytically write the single mode higher order covariance matrix after optical losses and from it derive the covariance matrix before losses. The matrix reads as following. As it contains long expressions, we present it column-wise:
\begin{equation}
\label{aloss}
    \gamma_\psi[:,1]=
\begin{pmatrix}
 t^2v(x)+r^2v(x_0)\\
 .\\
 .\\
 .\\
 .
 \end{pmatrix}
 \end{equation}
 \begin{equation*}
    \gamma_\psi[:,2]=
\begin{pmatrix}
 t^2c(x,p)\\
 t^2v(p) + r^2v(p_0)\\
 .\\
 .\\
 .
 \end{pmatrix}
 \end{equation*}
 \begin{equation*}
    \gamma_\psi[:,3]=
\begin{pmatrix}
t^3c(x^2,x) + 2tr^2\langle x\rangle v(x_0)  \\
 t^3c(p,x^2)\\
 t^4v(x^2) + 4t^2r^2\langle x_0^2 \rangle \langle x^2 \rangle + r^4v(x_0) \\
 .\\
 .
 \end{pmatrix}
 \end{equation*}
  \begin{equation*}
    \gamma_\psi[:,4]=
\begin{pmatrix}
t^3c(x,:xp:_W) + tr^22\langle p \rangle \langle x_0^2 \rangle   \\
t^3c(:xp:_W) + 2tr^2\langle x \rangle \langle p_0^2 \rangle\\
t^4c(x^2,:xp:_W)+2t^2r^2\langle x_0^2 \rangle \langle :xp:_W \rangle \\
\gamma[4,4]\\
.  

 \end{pmatrix}
 \end{equation*}
 \begin{equation*}
 \begin{split}
\gamma[4,4]&=t^4v(:xp:_W)+r^4v(:x_0p_0:_W)+\\&4t^2r^2(\langle p_0^2x^2 + x_0^2p^2 + p_0x_0xp + x_0p_0px\rangle)
\end{split}
 \end{equation*}
   \begin{equation*}
    \gamma_\psi[:,5]=
\begin{pmatrix}
 t^3 c(x,p^2)  \\
t^3c(p,p^2) + 2tr^2\langle p \rangle v(p_0)   \\
t^4c(x^2,p^2)+2t^2r^2\langle xpx_0p_0 + pxp_0x_0\rangle+r^4c(x_0^2,p_0^2)\\
t^4c(:xp:_W,p^2)+2t^2r^2\langle :xp:_W\rangle\langle p_0^2\rangle   \\
 t^4v(p^2) +4t^2r^2\langle p_0^2\rangle\langle p^2\rangle + r^4v(p_0^2)
 \end{pmatrix}.
 \end{equation*}

Here $::_W$ stands for Weyl-symmetric (with a different normalization $:xp:_W = xp+px$) and $v()$ for variance and $c()$ for covariance. The dots are inserted symbolically as the matrix is symmetric.
Together with equation
\begin{equation}
\begin{split}
   \expval{x}\rightarrow& t\expval{x}  \\
   \expval{p}\rightarrow& t\expval{p}\\
   \expval{xp+px} =& 2cov(x,p)+\expval{x}\expval{p}
\end{split}
\end{equation}
and \eqref{aloss}, the original higher order covariance matrix can be computed from the one measured after a lossy channel.

\subsection{Conversion between entries of the higher-order covariance matrix and moments of the rotated quadratures $X(\theta)$}\label{A:Dict}

Here, we briefly show, how the terms in the higher-order covariance matrix can be obtained from measurements of the generalized quadratures $X(\theta)$. Going by increasing level of nonlinearity:
\begin{equation}
    \begin{split}
        xp + px =& 2X(\frac{\pi}{4})^2 - x^2 - p^2\\
        x^2p + px^2 =& \frac{2}{3} [\sqrt{2}(X(\frac{\pi}{4})^3-X(-\frac{\pi}{4})^3)-p^3]\\
        xp^2 + p^2x =& \frac{2}{3} [\sqrt{2}(X(\frac{\pi}{4})^3+X(-\frac{\pi}{4})^3)-x^3]\\
        x^2p^2 + p^2x^2 =& \frac{1}{3}[2(X(\frac{\pi}{4})^4+X(-\frac{\pi}{4})^4)-3-x^4-p^4]
    \end{split}
\end{equation}

\begin{equation}
    \begin{split}
        xp^3 + p^3x =& \frac{1}{16\sqrt{3}}[16(3X(\frac{\pi}{3})^4 - X(\frac{\pi}{6})^4) \\&+ 6x^4 -26p^4 - 18(x^2p^2 + p^2x^2 + 1)]\\
        x^3p+px^3 =& \frac{1}{2(3-\frac{1}{3})\sqrt{3}}[16(X(\frac{\pi}{6})^4 - \frac{1}{3}X(\frac{\pi}{3})^4) \\&+ 2p^4 - (9-\frac{1}{3})x^4 - 6(x^2p^2 + p^2x^2 + 1)].
    \end{split}
\end{equation}
Therefore the higher-order covariance matrix $\gamma$ can be estimated from homodyne measurements under phase locks $\theta = -\frac{\pi}{4},0,\frac{\pi}{6},\frac{\pi}{4},\frac{\pi}{3},\frac{\pi}{2}$.

\subsection{Example: estimation of covariance from quadratures measured under three angles}\label{A:Estimation}
We can sample three variables $x,p$ and $y=\frac{1}{\sqrt{2}}(x+p)$ with different sample sizes $n_x$, $n_p$ and $n_y$. Given those, we can write the covariance of $x$ and $p$ as
\vspace{-1.05em}
\begin{equation}
        cov(x,p) = \frac{1}{2}E[2y^2 - x^2 - p^2] -E[x]E[p].
\end{equation}
With the sampled values, this can be rewritten as
\begin{equation}
    \begin{split}
        &cov(x,p)\\
        &\sim \frac{1}{2}(2\sum_i \frac{y_i^2}{n_y}-\sum_j \frac{x_j^2}{n_x}-\sum_k \frac{p_k^2}{n_p}) -\sum_l \frac{x_l}{n_x}\sum_m \frac{p_m}{n_p}\\
        &=cov(x,p) + [\bar{x}\bar{p}-\sum_l \frac{x_l}{n_x}\sum_m \frac{p_m}{n_p}].
    \end{split}
\end{equation}

\end{document}